\documentclass[12pt]{iopart}

\usepackage{graphicx}
\begin{document}

\title[Hydrodynamic Description of Heavy Ion Collisions]{
Hydrodynamic Description of Heavy Ion Collisions}

\author{Chiho Nonaka}

\address{Department of Physics, Nagoya University,  
Nagoya 464-8602, Japan}
\ead{nonaka@hken.phys.nagoya-u.ac.jp}
\begin{abstract}
We give a short review of hydrodynamic models at heavy ion collisions 
from the point of view of initial conditions, an equation of states (EoS)  
and freezeout process. Then we show our latest results of       
a combined fully three-dimensional 
macroscopic/microscopic transport approach. 
In this model for the early, dense, deconfined stage 
relativistic 3D-hydrodynamics of the reaction  
and a microscopic non-equilibrium model for the later hadronic
stage where the equilibrium assumptions are not valid anymore 
are employed.
Within this approach we study the dynamics of hot, bulk QCD matter, which
is being created in ultra-relativistic  heavy ion collisions
at RHIC. 
\end{abstract}

%\maketitle

\section{Hydrodynamic Models at RHIC}
The first five years of RHIC operations
at  $\sqrt{s_{NN}}=130$~GeV and $\sqrt{s_{NN}}=200$~GeV
have yielded a vast amount of interesting and
sometimes surprising results.  
There exists mounting evidence that RHIC has created
a hot and dense state of deconfined QCD matter with properties similar to
that of an ideal fluid \cite{Ludlam:2005gx,Gyulassy:2004zy} 
 -- this state of matter
has been termed the {\em strongly interacting Quark-Gluon-Plasma} (sQGP).
One of the evidence is the success of ideal hydrodynamic models in various physical  
observables.  Especially, for the first time, in elliptic flow the hydrodynamic limit 
shows good agreement with the experimental data at RHIC, though at AGS and 
SPS hydrodynamic models give larger value as compared to 
experimental data.    

The sophisticated 3D ideal hydrodynamic calculations, however, reveal 
that many of experimental data have not yet been fully evaluated 
or understood \cite{PCE}. 
For example, elliptic flow at forward and backward rapidity 
\footnote{Brazil group shows improved results of $v_2$ at 
forward/backward rapidity by using   
event-by-event fluctuated initial conditions \cite{Brasil06}.} 
and at peripheral collisions is overestimated by ideal hydrodynamical models.  
The Hanbury-Brown Twiss (HBT) puzzle is not completely understood  
from the point of view of hydrodynamics. 
These distinctions between hydrodynamic calculation and experimental data 
suggest that the ideal hydrodynamic picture is not applicable to all 
physical observables even in low transverse momentum region.     
At present our main interest in hydrodynamic models is the following: 
{\it How perfect the sQGP is?}  
      
Here, first, we shall give a short review of hydrodynamic models at heavy 
ion collisions.   
In addition to a numerical procedure for solving the relativistic 
hydrodynamic equation,  
hydrodynamic models are characterized by 
initial conditions, EoSs and freezeout process. 

The necessity of input of initial conditions for hydrodynamic models  
is one of the largest limitations of them. 
Because an initial condition is not be able to determined in the  
framework of hydrodynamic model itself, usually a parametrization 
of energy density and baryon number density based on    
Glauber type is used and parameters in it are determined by 
comparison with experimental data \cite{Kolb01, PCE, NoBa07}.   
Recently there are some studies in which more basic approaches, 
Color Glass Condensate (CGC) \cite{CGC_initial}, pQCD + saturation model 
\cite{Helsinki05} are used 
for construction of initial conditions.  

The most important advantage of hydrodynamic models is that it
directly incorporates an EoS as input and thus is so
far the only dynamical model in which a phase transition can
explicitly be incorporated. In the ideal fluid approximation 
-- and once an initial condition has been specified -- 
the EoS is the {\em only} input
to the equations of motion and relates directly to properties of the
matter under consideration. In this sense a hydrodynamic model is   
a bridge between QCD theory and experimental data and 
indispensable to describe heavy ion physics. 
However in usual practical hydrodynamic simulations, 
an EoS with 1st order phase transition (Bag model) is 
used. In fact, there are few studies on effect of order of QCD phase 
transition on physical observables \cite{Huovinen05}.    

Conventional hydrodynamic calculations need to assume a
{\em freezeout} temperature at which the hydrodynamic evolution is terminated
and a transition from the zero mean-free-path approximation of a
hydrodynamic approach to the infinite mean-free-path of free streaming
particles takes place. The freezeout temperature usually is a free
parameter which can be fitted to measured
hadron spectra. There are several approaches for dealing with 
freezeout process: chemical equilibrium \cite{Kolb01,Huovinen05}, 
partial chemical equilibrium \cite{PCE}, 
continuous emission model \cite{CEM} and construction of a 
hybrid model of  a hydro + cascade model \cite{CAS,NoBa07,Hirano06}. 

In Tab.~\ref{Tab-hydro} several hydrodynamic models are listed from 
the viewpoint of initial conditions and freezeout processes,  
because almost the same EoS with strong 1st order phase transition is used.   
Reference \cite{Kolb01} presents the first calculation which shows  
the remarkable agreement with experimental data for both of 
$P_T$ spectra and $v_2$ at RHIC.    
However it turns out that the assumption of 
single freezeout temperature   
where chemical freezeout and kinetic freezeout occur at the same time   
fails in reproducing hadron ratios correctly. 
To obtain correct proton $P_T$ spectra, we need to renormalize the 
$P_T$ spectra using the $p$ to $\pi$ ratio at the critical temperature.  
In addition, recently, Hirano and Gyullasy point out 
that the good agreement of elliptic flow with experimental data 
may accidentally happen. 
Hirano and  Kolb et al. propose that the introduction of two kinds 
of freezeout processes, chemical freezeout and kinetic freezeout to 
hydrodynamic models. 
In this model, normalization of the $P_T$ spectra for each particle  
are obtained correctly \cite{PCE},    
but to get better agreement with experiments in elliptic flow 
additional initial transverse flow is needed.     
At present, the combination of Gluaber type and a cascade model gives 
us the most promising result for both of $P_T$  spectra and $v_2$. 
Hirano et al.  perform calculations using CGC for initial condition and 
a cascade model for freezeout process, which suggests that viscosity is  
not negligible even at early stage of the expansion.   
In the next section we show our latest results \cite{NoBa07} based on 
the full 3D hydrodynamic approach \cite{Nonaka00} with the microscopic Ultra-relativistic
Quantum-Molecular-Dynamics (UrQMD) model \cite{uqmdref1}.    
%%%%%%%%%%%%%%%%%%%%%%%%%%%%%%%%%%%%%%%%%%%%%%%%%
%table for hydrodynamic models 
\begin{table}[h]
\begin{center}
\caption{Hydrodynamic models at RHIC.}
\label{Tab-hydro}
\begin{tabular}{c|l|l}
\hline
Reference  & Initial Conditions & Freezeout Process \\ \hline \hline
\cite{Kolb01}      & Glauber type  &  chemical equilibrium   \\ \hline
\cite{PCE}      &  Glauber type & partial chemical equilibrium \\ \hline
\cite{CAS,NoBa07}      &  Glauber type & cascade model      \\ \hline
\cite{HiNa04}      &  CGC & partial chemical equilibrium   \\ \hline
\cite{Hirano06}      &  CGC & cascade model  \\ \hline
\end{tabular}
\end{center}
\end{table}
%%%%%%%%%%%%%%%%%%%%%%%%%%%%%%%%%%%%%%%%%%%%%%%%%

\section{Hydro+UrQMD Model}
We calculate hadron distribution at switching temperature 
from the 3D hydrodynamic model using Cooper-Frye formula \cite{CF} 
and produce initial conditions for UrQMD model by Monte Carlo from it. 
Such hybrid macro/micro transport calculations  are to date the
most successful approaches
for describing the soft physics at RHIC. The biggest advantage of the
hydrodynamic description  is
that it directly incorporates an EoS as input - one of its
largest limitations is that it requires thermalized initial conditions
and one is not able to do an ab-initio calculation.

\begin{figure}[h]
\includegraphics[width=0.90\linewidth]{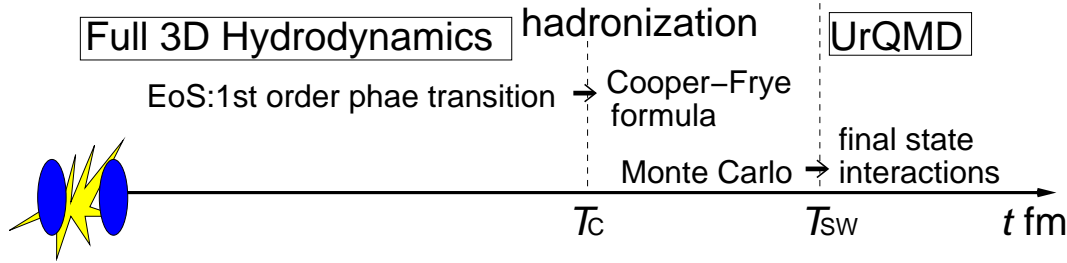}
\caption{Schematic sketch of 3D hydro+UrQMD model. 
$T_{\rm c}(=160$ MeV)  and $T_{\rm SW}(= 150)$ MeV are critical temperature and 
switching temperature from hydrodynamics to UrQMD model, respectively. 
\label{Fig-3dhu}
}
\end{figure}

Figure \ref{Fig-3dhu} shows a schematic sketch of the full 3D 
hydrodynamic model + UrQMD. 
After heavy ion collisions, first, hydrodynamic expansion starts. 
We introduce one more parameter, switching temperature 
from hydrodynamic picture to hadron base event generator, UrQMD. 
This switching temperature should be  just below the critical 
temperature. Here it is set to 160 MeV. 

%3d hydro + urqmd
%%%%%%%%%%%%%%%%%%%%%%%%%%%%%%%%%%%%%%%%%%%%%%%%%%%%%%%%%%
%%%%%%%%%%%%%%%%%%%%%%%%%%%%%%%%%%%%%%%%%%%%%%%%%%%%%%%%%%

% Pt spectra
Figure 2 shows the $P_T$ spectra of $\pi^+$, $K$ and $p$ at 
$\sqrt{s_{NN}}=200$ GeV central collisions. The most compelling feature 
is that the hydro+micro approach is capable of accounting for the proper 
normalization of the spectra for all hadron species
 without any additional correction as is performed in the pure hydrodynamic model. 
The introduction of a realistic freezeout process  
provides therefore a natural solution to the problem of separating chemical and 
kinetic freeze-out in a pure hydrodynamic approach. 

%centrality dependence of Pt spectra
In Fig.~3 centrality dependence of $P_T$ spectra of 
$\pi^+$ is shown. The impact parameter for each centrality is determined 
simply by the collision geometry.  
The separation between model results and experiment appears  
at lower transverse momentum in peripheral collisions compared to central 
collisions, just as in the pure hydrodynamic calculation.
The 3D hydro + micro model does not provide any improvement for this behavior, 
since the hard physics high $P_T$  contribution to the spectra
occurs at early reaction times before the system has reached the QGP phase
and is therefore neither included in the pure 3D hydrodynamic calculation 
nor in the hydro+micro approach.

%%%%%%%%%%%%%%%%%%%%%%%%%%%%%%%%%%%%%%%%%%%%%%%%%%%%%%%%%%%%
\begin{figure}[tbh]
\begin{minipage}[t]{80mm}
%Pt spectra 2
\includegraphics[width=1.0\linewidth]{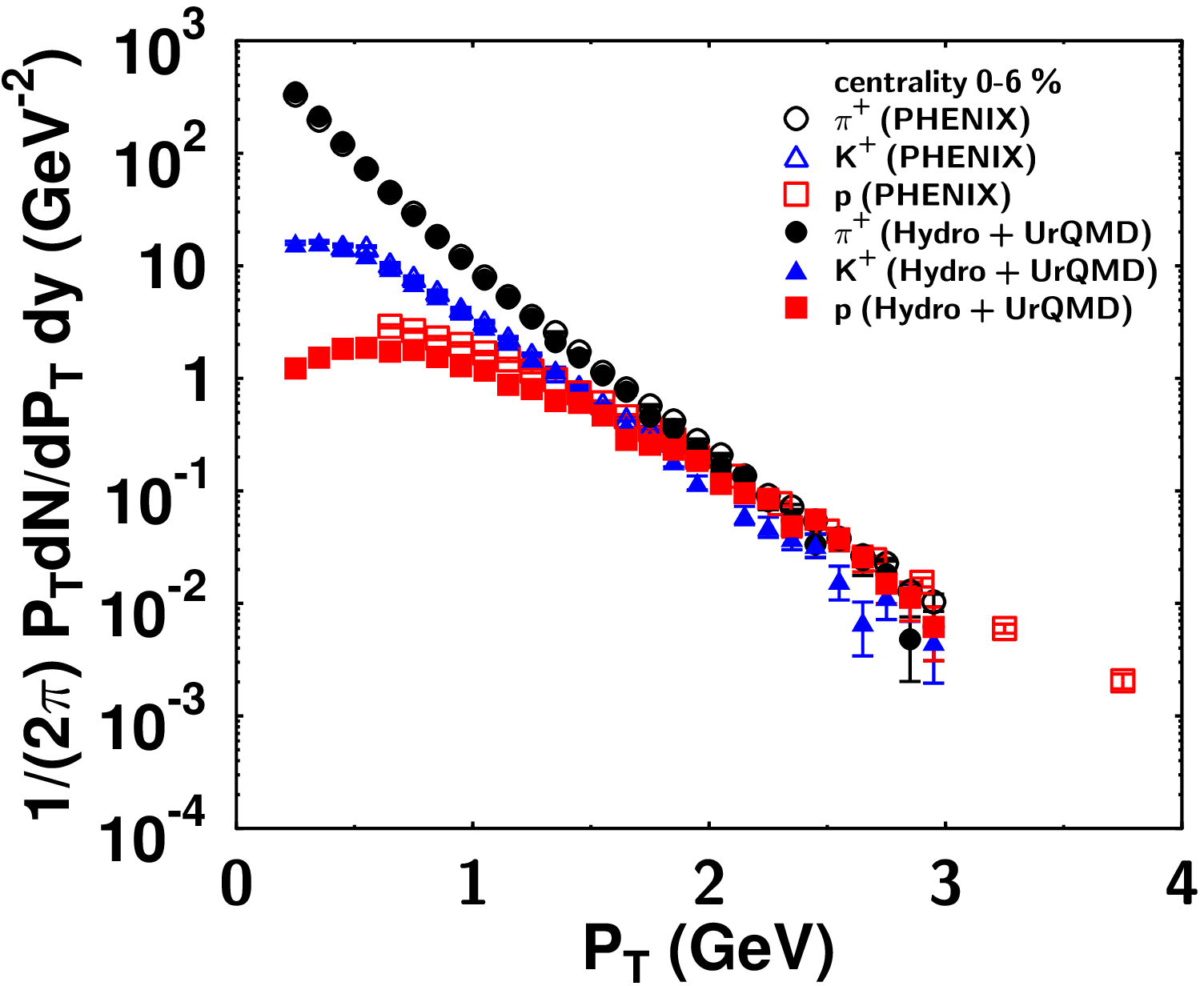}
\caption{$P_T$ spectra for $\pi^+$, $K^+$ and $p$ at central collisions 
with PHENIX data \cite{PHENIX_PT}. 
}
\end{minipage}
\hspace{1mm}
\begin{minipage}[t]{80mm}
%Centrality dependence of Pt spectra 3 
\includegraphics[width=1.0\linewidth]{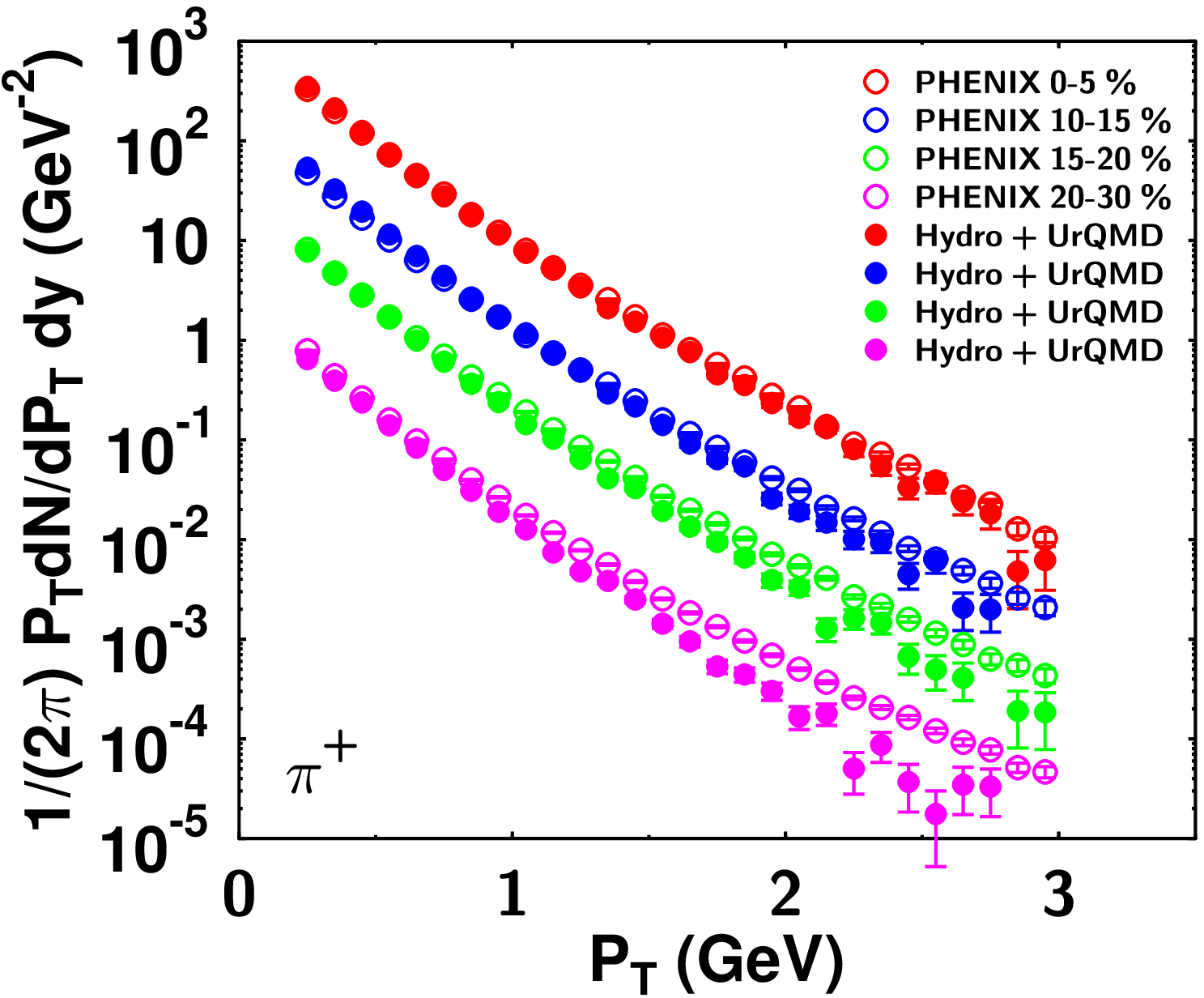}
\caption{Centrality dependence of $P_T$ spectra of $\pi^+$ with
PHENIX data \cite{PHENIX_PT}. The $P_T$ spectra at 10--15 \%, 15--20
\% and  20--30 \% are divided by 5, 25 and 200, respectively.
}
\end{minipage}
\end{figure}
%%%%%%%%%%%%%%%%%%%%%%%%%%%%%%%%%%%%%%%%%%%%%%%%%%%%%%%%%%%%

%centrality dependence of the pseudorapidity 
Figure 4 shows the centrality dependence of the  
pseudorapidity distribution of charged hadrons compared to 
PHOBOS data \cite{PHOBOS_eta}.
Solid circles stand for model results and open circles denote
data taken by the PHOBOS collaboration \cite{PHOBOS_eta}.
The impact parameters are set to $b=2.4, 4.5, 6.3, 7.9$ fm for 
0-6 \%, 6-15 \%, 15-25 \% and 25-35 \% centralities, respectively. 
Our results are consistent with experimental data over a wide pseudorapidity 
region. 
There is no distinct difference between 3-D ideal
hydrodynamic model and the hydro + UrQMD model in the centrality dependence of the 
psuedorapidity distribution, indicating that the shape of psuedorapidity 
distribution is insensitive to the detailed microscopic 
reaction dynamics of the hadronic final state \cite{NoBa07}.

%pt spectra for multi-strange particles
In Fig.~5 we analyze the $P_T$ spectra of 
multistrange particles. Our results show good agreement with   
experimental data for $\Lambda$, $\Xi$, $\Omega$ for centralities 0--5 \%.  
In this calculation the additional procedure for normalization 
is not needed (Fig.~6).  
Recent experimental results suggest that 
at thermal freezeout multistrange baryons exhibit less transverse flow  
and a higher temperature closer to the chemical freezeout 
temperature compared to non- or single-strange baryons
\cite{STAR_strange1,STAR_strange2}. This behavior can be understood in terms
of the flavor dependence of the hadronic cross section, which decreases
with increasing strangeness content of the hadron. The reduced
cross section of multi-strange baryons leads to a decoupling from the
hadronic medium at an earlier stage of the reaction, allowing them
to provide information on the properties of the hadronizing QGP less
distorted by hadronic final state interactions 

%%%%%%%%%%%%%%%%%%%%%%%%%%%%%%%%%%%%%%%%%%%%%%%%%%%%%%%%%%%%
\begin{figure}[tbh]
\begin{minipage}[t]{80mm}
%eta distribution, central dependence 3 
\includegraphics[width=1.0\linewidth]{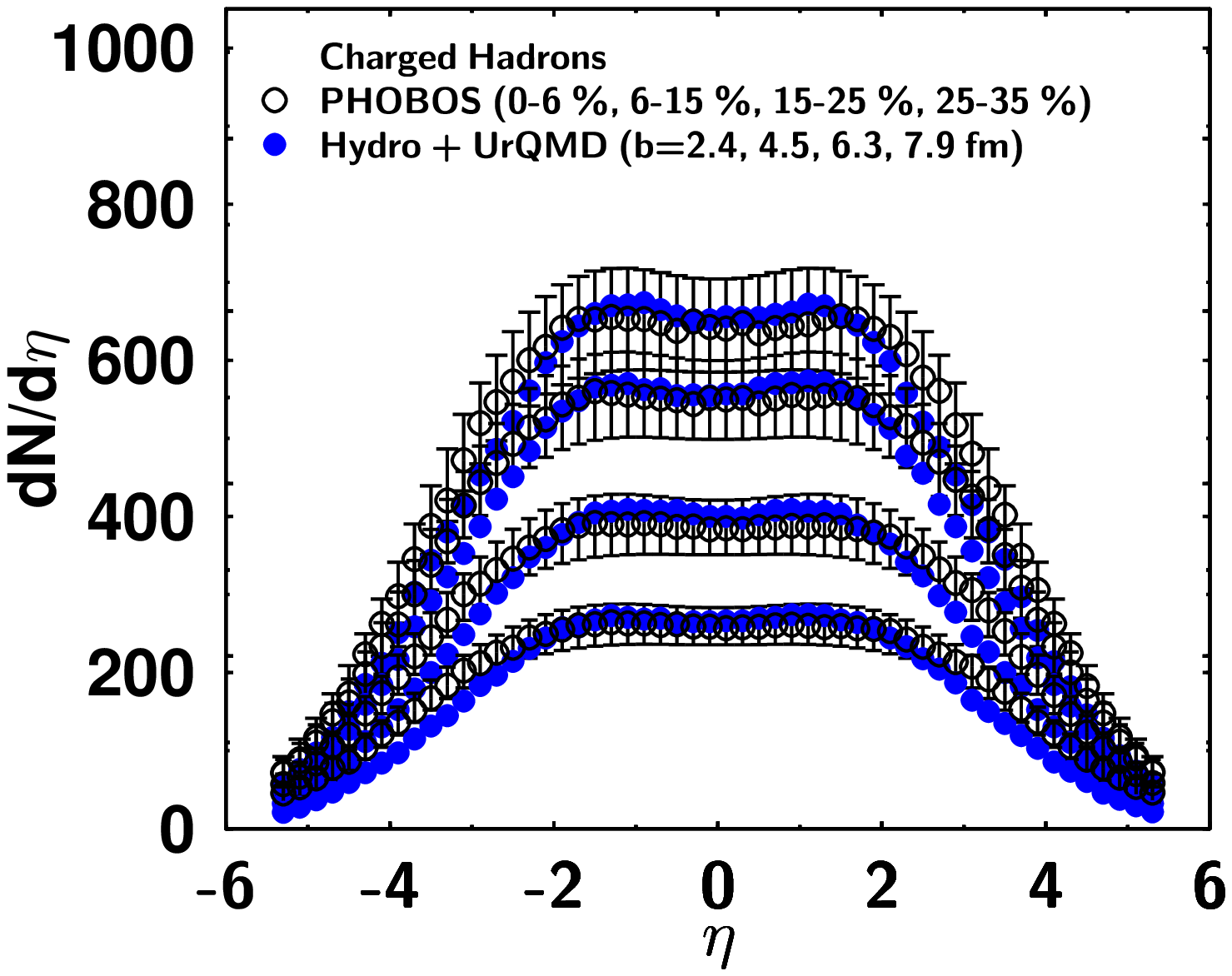}
\caption{Centrality dependence of pseudorapidity distribution of
charged particles with PHOBOS data \cite{PHOBOS_eta}.
}
\end{minipage}
\hspace{1mm}
\begin{minipage}[t]{80mm}
\includegraphics[width=1.0\linewidth]{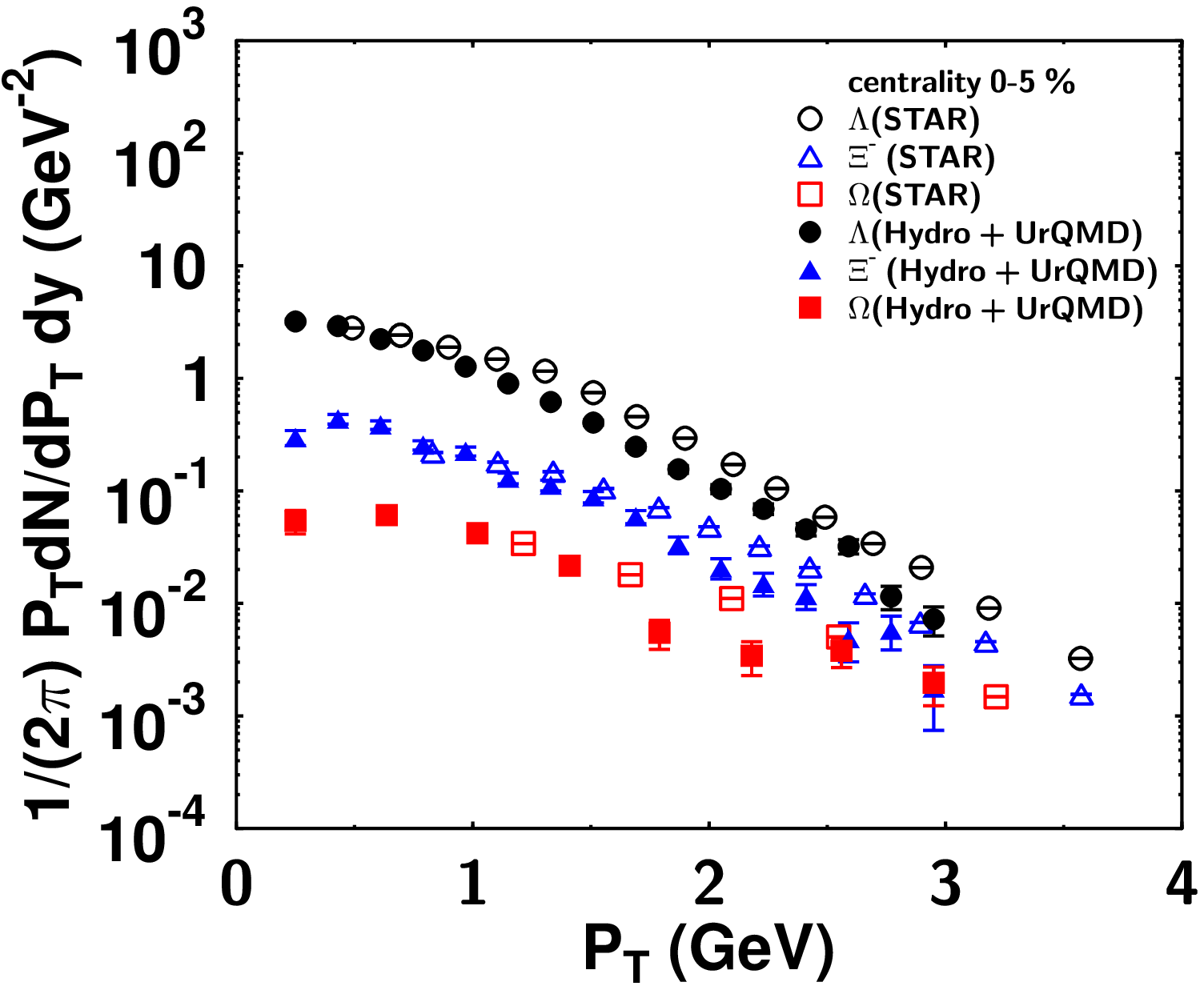} %4
\caption{$P_T$ spectra of multi-strange particles at centralities 0--5 \%  
and 10--20 \% with STAR data \cite{STAR_strange1}.
}
\end{minipage}
\label{Fig-hydro_v2_pt}
\end{figure}
%%%%%%%%%%%%%%%%%%%%%%%%%%%%%%%%%%%%%%%%%%%%%%%%%%%%%%%%%%%%

%meanpt as a function of Pt
In Fig.~7 the mean transverse momentum $\langle P_T\rangle $ 
as a function of hadron  
mass  is shown. Open symbols denote the value at $T_{\rm sw}=150$~MeV, corrected
for hadronic decays. Not surprisingly, in 
this case the $\langle P_T\rangle $   follow a straight line, suggesting a 
hydrodynamic expansion.  However if hadronic rescattering 
is taken into account (solid circles) the $\langle P_T\rangle $  
do not follow the straight line any more: 
the $\langle P_T\rangle $ of pions is actually reduced by hadronic
rescattering (they act as a heat-bath in the collective expansion), 
whereas protons actually pick up additional transverse momentum in the
hadronic phase. RHIC data by the STAR collaboration is shown via
the solid triangles -- overall the proper treatment of hadronic
final state interactions significantly improves the agreement of the
model calculation with the data.

%%%%%%%%%%%%%%%%%%%%%%%%%%%%%%%%%%%%%%%%%%%%%%%%%%%%%%%%%%%%
\begin{figure}[tbh]
\begin{minipage}[t]{80mm}
\includegraphics[width=1.0\linewidth]{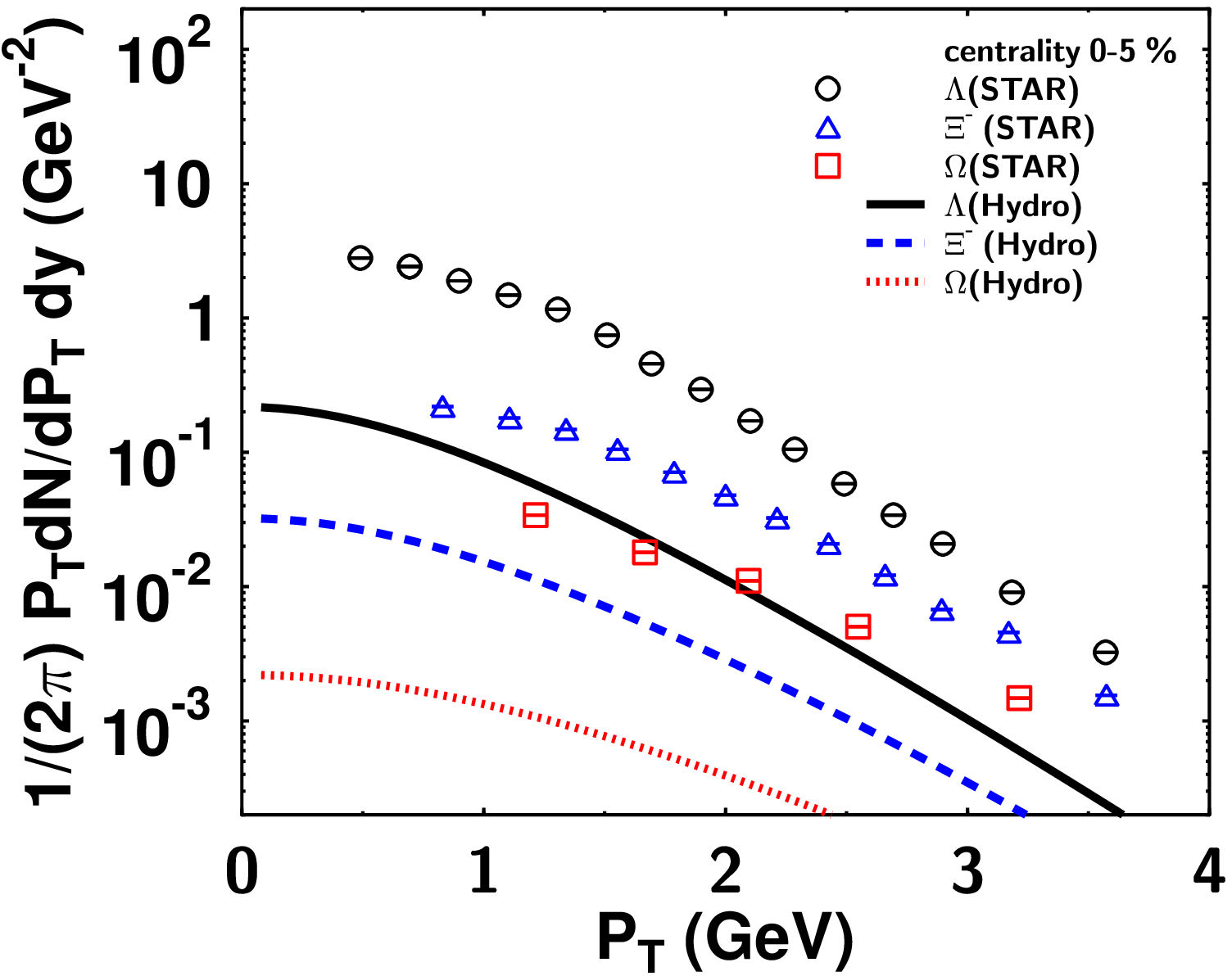} %6
\caption{$P_T$ spectra for multistrange baryons at central collisions 
from pure hydro with STAR data \cite{STAR_strange1}. 
}
\end{minipage}
\hspace{1mm}
\begin{minipage}[t]{80mm}
\includegraphics[width=1.0\linewidth]{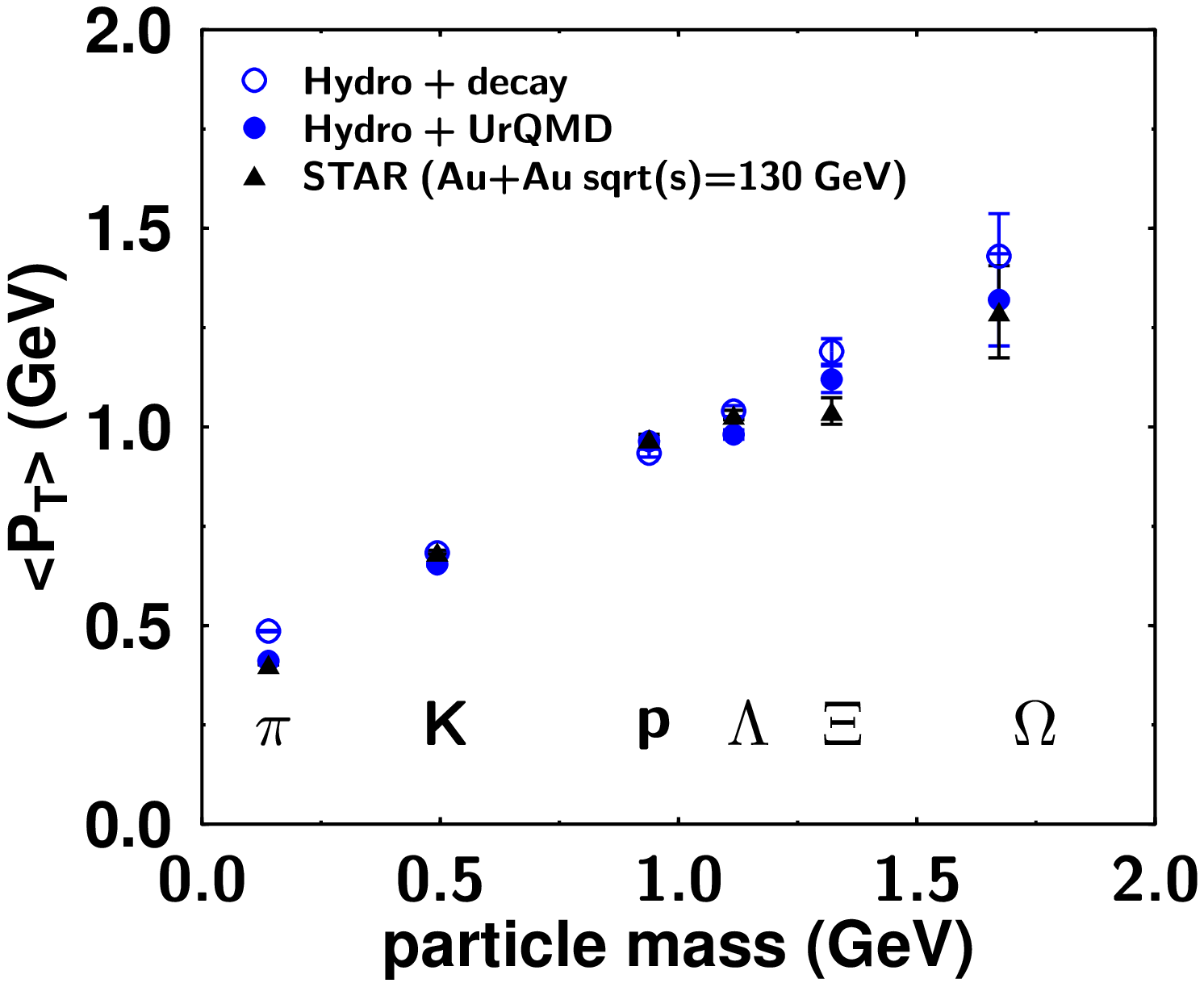} %7
\caption{Mean $P_T$ as a function of mass with  
STAR data (Au+Au $\sqrt{s_{NN}}=130$ GeV) \cite{STAR_strange2}.
}
\end{minipage}
\end{figure}
%%%%%%%%%%%%%%%%%%%%%%%%%%%%%%%%%%%%%%%%%%%%%%%%%%%%%%%%%%%%

% v2 as a function of Pt, 
In Fig.~8 we plot elliptic flow $v_2$ as a function of $P_T$. 
The solid line stands for the pure hydro calculation, terminated at the switching 
temperature $T_{\rm sw}$ and solid circles denote the full
hydro+micro calculation. We find that the QGP contribution to the 
elliptic flow depends on the transverse momentum -- for low $P_T$ nearly
100\% of the elliptic flow is created in the QGP phase of the reaction,
whereas the hadronic phase contribution increases to 25\% at a $P_T$ 
of 1~GeV/$c$. 

%v2 as a function of eta, decay, final states effect 
Figure 9 shows the elliptic flow as a 
function of $\eta$: the pure hydrodynamic calculation is shown by 
the solid curve, the hydrodynamic contribution at $T_{\rm sw}$ is denoted
by the dashed line and the full hydro+micro calculation is given by
the solid circles, together with PHOBOS data (solid triangles). 
The shape of  the elliptic flow in the pure hydrodynamic calculation at 
$T_{\rm sw}$ is quite different from that of the full
hydrodynamic one terminated at a freeze-out temperature of 110~MeV. 
Apparently the slight bump at forward and backward rapidities observed
in the full hydrodynamic calculation develops first in the later hadronic
phase, since it is not observed in the calculation terminated at 
$T_{\rm sw}$.
Evolving the hadronic phase in the hydro+micro approach will increase
the elliptic flow at central rapidities, but not in the projectile and 
target rapidity domains.
As a result, the elliptic flow calculation in the hydro+micro approach
is closer to the experimental data when compared to the pure hydrodynamic calculation.
%%%%%%%%%%%%%%%%%%%%%%%%%%%%%%%%%%%%%%%%%%%%%%%%%%%%%%%%%%%%
\begin{figure}[tbh]
\begin{minipage}[t]{80mm}
\includegraphics[width=1.0\linewidth]{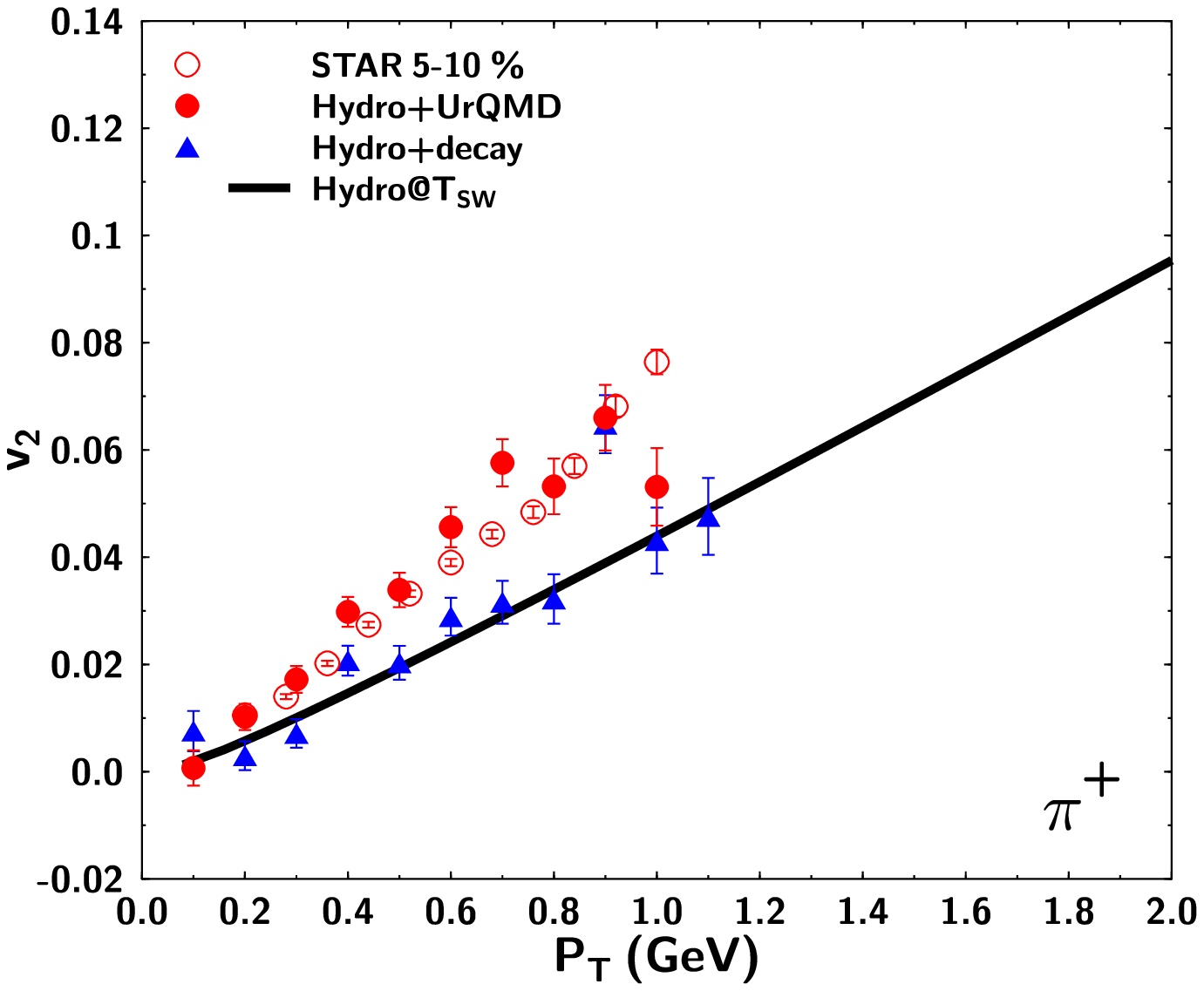} % 7
\caption{Elliptic flow as a function of $P_T$ of $\pi^+$ at 
centrality 5-10 \% with STAR data \cite{STAR_v2}. 
}
\end{minipage}
\hspace{1mm}
\begin{minipage}[t]{80mm}
\includegraphics[width=1.0\linewidth]{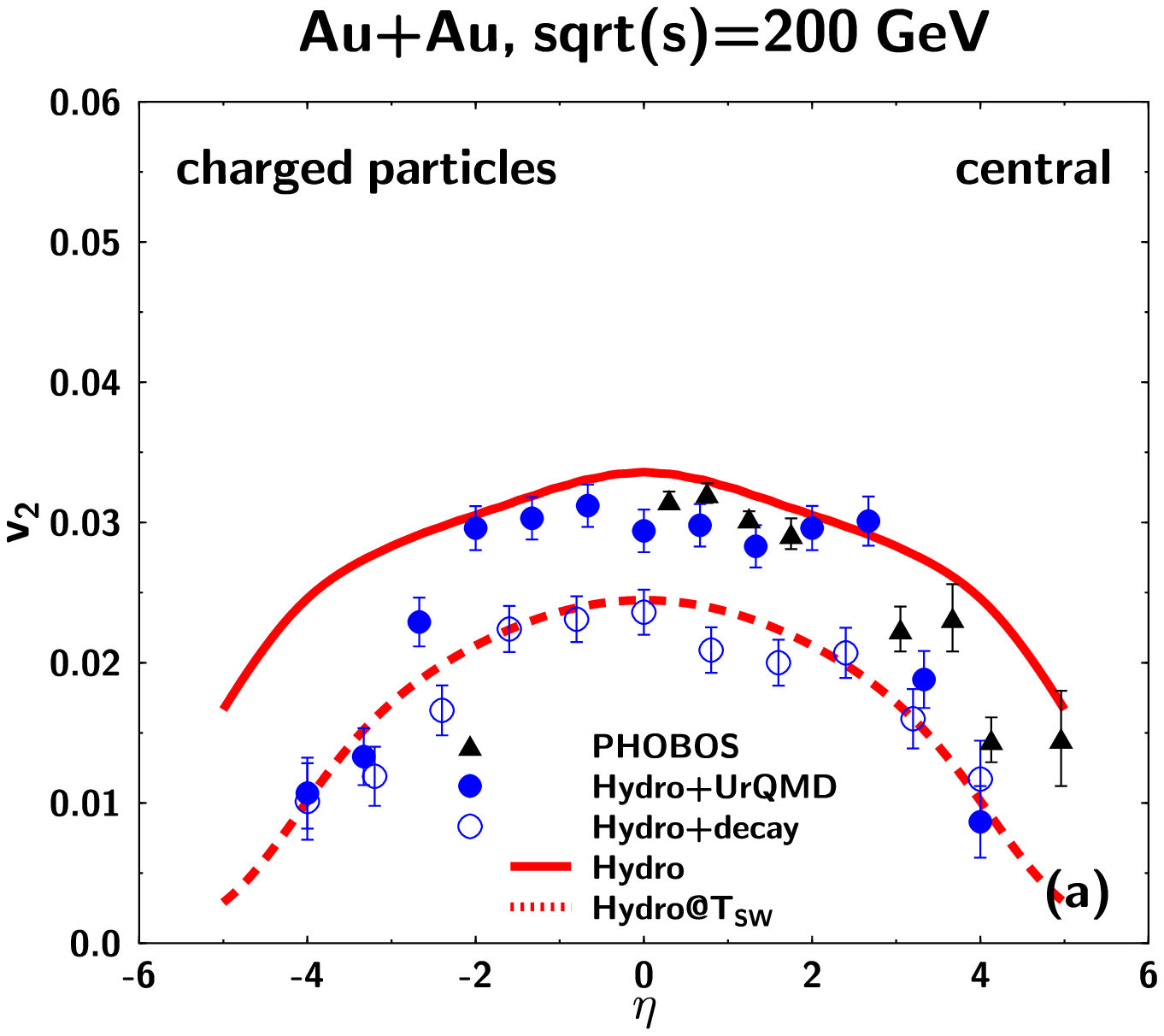}  % 8
\caption{Elliptic flow as a function of $\eta$ of charged particles 
with PHOBOS data \cite{PHOBOS_v2_eta}. 
}
\end{minipage}
\label{Fig-hydro_v2_pt}
\end{figure}
%%%%%%%%%%%%%%%%%%%%%%%%%%%%%%%%%%%%%%%%%%%%%%%%%%%%%%%%%%%%

We developed a novel implementation of 
the well known  hybrid macroscopic/microscopic transport approach,
combining a newly developed relativistic 3+1 dimensional
hydrodynamic model for the early deconfined stage of the reaction 
and the hadronization process
with a microscopic non-equilibrium model for the later hadronic
stage.  

Within this approach we have dynamically 
calculated the
freezeout of the hadronic system,
accounting for the collective flow on the
hadronization hypersurface generated by the QGP expansion.
We have compared the results of our hybrid model and of a calculation 
utilizing our hydrodynamic model for the full evolution of the reaction to
experimental data. This comparison has allowed us to quantify the strength
of dissipative effects prevalent in the later hadronic phase of the reaction,
which cannot be properly treated in the framework of ideal hydrodynamics.

\section{Summary}
The full 3D relativistic hydrodynamics + cascade model is one of successful 
and realistic models for description of dynamics of hot QCD bulk matter at RHIC, 
which helps us to understand medium property at RHIC in detail.   
Using this model,  we can explore interesting phenomena which are caused by 
interactions between medium and jets \cite{Bass, Renk}.  
One of proposed interesting physical observables is 
mach cone \cite{Renk} for the wake of them, from which we can know medium property 
in detail, too.  

However still further investigation for EoS and initial conditions in 
hydrodynamics is needed.  
Especially the EoS is the key to know the QCD phase transition directly from 
comparison with experimental data.      
Recent lattice QCD calculation shows relatively high critical 
temperature with crossover phase transition and the existence of critical 
end point \cite{Hatsuda}, which suggests that the EoS with strong 1st phase 
transition is not realistic. 
Becuase in hydrodynamic calculations, however, outputs may also be changed 
easily by choice of initial conditions and freezeout process, 
which makes it difficult to obtain the conclusive discussion on the EoS.  
First of all, we have to determine the most realistic initial conditions and 
freezeout process in hydrodynamic models before a detailed discussion about the EoS.  

Furthermore, viscosity effect in medium also starts to be discussed 
actively \cite{viscous_hydro}. 
Due to serious difficulty in construction not only 
of a viscous hydrodynamic code but also of framework of viscous hydrodynamics 
without the causality problem, the progress of study of viscous hydrodynamics 
has been slow.     
However, by virtue of recent rapid development,      
the practical calculations with viscous hydrodynamics which are comparable to 
experimental data  will be achieved in the near future \cite{viscous_hydro}.            

Finally, at LHC heavy ion collisions  experiments will start in a year.   
Upcoming LHC data shall bring a lot of interesting, fruitful and even unexpected  
results for QGP physics, where hydrodynamic description will be    
helpful and useful for understanding of it.

\vspace{-5mm}
\section*{Acknowledgment}
We would like to thank Steffen A. Bass, Berndt Muller, Masayuki Asakawa and Joseph 
Kapusta for many valuable discussions and encouragement in pursuit of this work. 
The work is supported by Saneyoshi Shougakukai and 
the 21st century COE ``The Origin of the Universe and Matter: Physical Elucidation 
of the Cosmic History'' program of Nagoya University.


\vspace{-5mm}
\section*{References}
\begin{thebibliography}{99}

\bibitem{Ludlam:2005gx} 
T.~Ludlam,           
%``Experimental results from the early measurements at RHIC: Hunting 
%for the quark-gluon plasma,''                                       
Nucl.\ Phys.\ A {\bf 750}, 9 (2005). 
%%CITATION = NUPHA,A750,9;%%     
\bibitem{Gyulassy:2004zy} 
M.~Gyulassy and L.~McLerran, 
%``New forms of QCD matter discovered at RHIC,'' 
Nucl.\ Phys.\ A {\bf 750}, 30 (2005).              
%[arXiv:nucl-th/0405013].             
%%CITATION = NUCL-TH 0405013;%% 

\bibitem{Brasil06}
%Examining the Necessity to Include Event-By-Event Fluctuations in Experimental 
%Evaluations of Elliptical Flow
%R.~Andrade, F.~Grassi, Y.~Hama, T.~Kodama and O. Socolowski, Jr.,
R.~Andrade et al., Phys.\ Rev.\ Lett.\ {\bf 97}, 202302 (2006).  

\bibitem{Kolb01}
%Elliptic flow at SPS and RHIC: from kinetic transport to hydrodynamics
P.~F.~Kolb, P.~Huovinen, U.~Heinz, H.~Heselberg,
Phys.\ Lett.\ {\bf B500}, 232 (2001);
%Radial and eliptic flow at RHIC:further predictions,
P.~Huovinen, P.~F.~Kolb, U.~Heinz, P.~V.~Ruusukanen, S.~A.~Voloshin,
Phys.\ Lett. \ {\bf B503}, 58 (2001).

\bibitem{PCE}
%Collective flow and two pion correlations from a relativistic
%hydrodynamic model with early chemical freezeout.
T.~Hirano and K.~Tsuda, Phys.\ Rev.\ {\bf C66}, 054905 (2002);
%Transverse flow and hadrochemistry in Au+Au collisions
%at (S(NN))**(1/2) = 200-GeV
P.~F.~Kolb, R.~Rapp, Phys.\ Rev.\ {\bf C67}, 044903 (2003).

\bibitem{CGC_initial}
T.~Hirano, these proceedings; R.~Fries, these proceedings. 

\bibitem{Helsinki05}
%HIC-tested predictions for low-p(T) and high-p(T) hadron spectra in 
%nearly central Pb + Pb collisions at the LHC
%K.~J.~Eskola, H.~Honkanen, H.~Niemi, P.~V.~Ruuskanen, S.~S.~Rasanen, 
K.~J.~Eskola et al., Phys.\ Rev.\ {\bf C72}, 044904 (2005). 

\bibitem{Huovinen05}
%Anisotropy of flow and the order of phase transition in 
%relativistic heavy ion collisions.
P.~Huovinen, Nucl.\ Phys.\ {\bf A761}, 29 (2005). 

\bibitem{CEM}
F.~Grassi, Y.~Hama and T.~Kodama, Phys.\ Lett.\ {\bf 355}, 
9 (1995); Z.\ Phys.\ {\bf C73}, 153 (1996). 

\bibitem{CAS}
%Dynamics of hot bulk QCD matter: From the quark gluon 
%plasma to hadronic freezeout
S.~A.~Bass, A.~Dumitru, Phys.\ Rev.\ {\bf C61}, 064909 (2000); 
%Flow at the SPS and RHIC as a quark gluon plasma signature
D.~Teaney, J.~Lauret,E.~V.~Shuryak, 
Phys.\ Rev.\ Lett.\ {\bf 86}, 4783 (2001); nucl-th/0110037.   

\bibitem{NoBa07}
C.~Nonaka, S.~A.~Bass, Phys.\ Rev.\ {\bf C75} 014902 (2007).


\bibitem{Hirano06}
%Hadronic dissipative effects on elliptic flow in ultrarelativistic 
%heavy-ion collisions
T.~Hirano, U.~Heinz, D.~Kharzeev, R.~Lacey, Y.~Nara, 
Phys.\ Lett.\ {\bf B636}, 299 (2006). 

\bibitem{Nonaka00}
%(3+1)-dimensional relativistic hydrodynamical expansion of 
%hot and dense matter in ultrarelativistic nuclear collision
C.~Nonaka, E.~Honda, S.~Muroya, Eur.\ Phys.\ J. {\bf C17}, 663 (2000).

\bibitem{uqmdref1}
        S.~A.~Bass et al.,  Progr. Part. Nucl. Physics Vol. {\bf 41},~225~(1998)
        %M.~Belkacem, M.~Bleicher, M.~Brandstetter, L.~Bravina,
        %C.~Ernst, L.~Gerland, M.~Hofmann, S.~Hofmann, J.~Konopka, G.~Mao, 
        %L.~Neise, S.~Soff, C.~Spieles, H.~Weber, L.~A.~Winckelmann, 
        %H.~St\"ocker, W.~Greiner, C.~Hartnack, J.~Aichelin and N.~Amelin, 
        %\newblock 
        Progr. Part. Nucl. Physics Vol. {\bf 41},~225~(1998); 
        %nucl-th/9803035\\
        M. Bleicher et al., 
        %E.~Zabrodin, C.~Spieles, S.A.~Bass, C.~Ernst, 
        %S.~Soff, L.~Bravina, M.~Belkacem, H.~Weber, H.~St\"ocker 
        %and W.~Greiner, 
        J. Phys. {\bf G25},~1859~(1999). 
%        The UrQMD source code is available via the URL:\\
%        \verb[http://www.th.physik.uni-frankfurt.de/~urqmd[

\bibitem{HiNa04}
%Hydrodynamic afterburner for the color glass
%condensate and the parton energy loss
T.~Hirano, Y.~Nara, Nucl.\ Phys.\ {\bf A743}, 305 (2004).

\bibitem{CF}
        F.~Cooper and G.~Frye,
       Phys.~Rev.~{\bf D10}, 186~(1974).

\bibitem{PHENIX_PT}
S.~S.~Adler et al. (PHENIX Collaboration), Phys.\ Rev.\ {\bf C69},  
034909 (2004). 

\bibitem{PHOBOS_eta}
B.~B.~Back et al. (PHOBOS Collaboration), Phys.\ Rev.\ Lett.\ {\bf 91},   
052303 (2003). 

\bibitem{STAR_strange1}
%strange baryon production in Au+Au collisions at
%top RHIC energy as a probe of bulk properties
M.~Estienne (for the STAR Collaboration), J.\ Phys.\ {\bf G31}, S873
(2005).
%nucl-ex/0412041

\bibitem{STAR_strange2}
%Multistrange Baryon Production in Au-Au Colisions at 
%\sqrt{S_NN} = 130 GeV.
J.~Adames et al. (STAR Collaboration), Phys.\ Rev.\ Lett.\ {\bf 92}, 
182301 (2004). 

\bibitem{STAR_v2}
% Azimuthal anisotropy in Au+Au collisions at $\sqrt{s_{NN}}=200$ GeV
%nucl-ex/0409033
J.~Adams et al. (STAR Collaboration), Phys.\ Rev.\ {\bf C72}, 
014904 (2005).

\bibitem{PHOBOS_v2_eta}
%Centrality and pseudorapidity dependence of elliptic flow for charged 
%hadrons in Au+Au collisions at sqrt(sNN) = 200 GeV
B.~B.~Back et al. (PHOBOS Collaboration), Phys.\ Rev.\ {\bf C72},  
051901(R) (2005).

\bibitem{Bass}
%Jet-quenching in a 3D hydrodynamic medium
T.~Renk, J.~Ruppert, C.~Nonaka, S.~A.~Bass, nucl-th/0611027, to appear in 
Phys.\ Rev.\ C.  

\bibitem{Renk}
%Mach Cone in quark-gluon plasma
Casalderrey-Solana, these proceedings;   
%Mach cones and dijets -angular correlations as a probe of early medium evolution, 
T.~Renk, these proceedings.  

\bibitem{Hatsuda}
T.~Hatsuda, these proceedings. 

\bibitem{viscous_hydro}
%Experimental evidence of perfet ffluidity at RHIC
D.~Teaney, these proceedings;
%New Formulation of Causal Dissipative Hydrodynamics: Shock wave propagation.
%Ph.~Mota, G.~S.~Denicol, T.~Koide, T.~Kodama, these proceedings;
T.~Kodama, these proceedings;
% Relativistic Dynamics of Non-ideal Fluids: Viscous and heat-conducting
%fluids. II. Transport properties and microscopic description of relativistic
%nuclear matter
A.~Muronga, nucl-th/0611090, nucl-th/0611091;
%Dissipative hydrodynamics for viscous relativistic fluids
U.~Heinz, H.~Song, A.~K.~Chaudhuri, Phys.\ Rev.\ {\bf C73},034904,2006.


\end{thebibliography}
\end{document}